\documentclass[%
 reprint,
superscriptaddress,
showpacs,
 amsmath,amssymb,
 aps,
floatfix,
]{revtex4-1}

\usepackage{dcolumn}
\usepackage{bm}

\usepackage[dvipdfmx]{graphicx,color}
\usepackage{color}

\usepackage{ulem} 

\begin{document}


\title{Incident-energy-dependent spectral weight of resonant inelastic x-ray scattering in doped cuprates}

\author{Kenji Tsutsui}
\email{tutui@spring8.or.jp}
\affiliation{Synchrotron Radiation Research Center, National Institutes for Quantum and Radiological Science and Technology, Hyogo 679-5148, Japan}

\author{Takami Tohyama}
\email{tohyama@rs.tus.ac.jp}
\affiliation{Department of Applied Physics, Tokyo University of Science, Tokyo 125-8585, Japan}

\date{\today}
             
\pacs{78.70.Ck, 78.20.Bh, 74.72.-h}

\begin{abstract}

We theoretically investigate the incident-photon energy $\omega_\mathrm{i}$ dependence of resonant inelastic x-ray scattering (RIXS) tuned for the Cu $L$ edge in cuprate superconductors by using the exact diagonalization technique for a single-band Hubbard model.
Depending on the value of core-hole Coulomb interaction in the intermediate state, RIXS for non-spin-flip channel shows either a $\omega_\mathrm{i}$-dependent fluorescencelike or $\omega_\mathrm{i}$-independent Raman-like behavior for hole doping.
An analysis of x-ray absorption suggests that the core-hole Coulomb interaction is larger than on-site Coulomb interaction in the Hubbard model, resulting in a fluorescence-like behavior in RIXS consistent with recent RIXS experiments.
A shift on the high-energy side of the center of spectral distribution is also predicted for electron-doped systems though spectral weight is small.
Main structures in the spin-flip channel exhibit a Raman-like behavior as expected, accompanied with a fluorescencelike behavior with small intensity.

\end{abstract}
\maketitle


\section{Introduction}

Resonant inelastic x-ray scattering (RIXS) experiments tuned for the Cu $L$ edge have provided a lot of new insights about spin dynamics of the spin-flip channel in cuprate superconductors when incident photon has the $\pi$ polarization~\cite{Ament2011,Dean2015}.
The non-spin-flip channel involving charge dynamics as well as two-magnon excitations can be predominantly detected by $\sigma$ polarization for the incident photon. 

Not only polarization dependence but also incident-photon energy $\omega_\mathrm{i}$ dependence of the RIXS spectrum gives us useful information on the electronic states of cuprates.
Experimentally it has been reported that the $\omega_\mathrm{i}$ dependence for the $\pi$-polarized incident photon gives a Raman-like behavior independent of $\omega_\mathrm{i}$, while the $\sigma$ polarization induces a fluorescencelike shift of spectral weight with increasing $\omega_\mathrm{i}$~\cite{Minola2015,Huang2016}.
Theoretically it has been pointed out that the $\omega_\mathrm{i}$ dependence is sensitive to particle-hole excitations of quasiparticles~\cite{Okada2000,Benjamin2014} depending on band structures of materials~\cite{Kanasz-Nagy2015}.
On the other hand, a model calculation for the doped-Mott insulator, i.e., a calculation of the single-band Hubbard model, has also captured Raman-like and fluorescencelike behaviors~\cite{Huang2016}:  The Raman-like excitation comes from collective spin excitations, while the fluorescencelike shift is due to the continuum of particle-hole excitations. 

In the fluorescence of the normal x-ray emission spectroscopy (XES), there is no correlation between a photoexcited electron in the vacuum continuum and a valence electron decaying into the core hole so that the emission spectrum is almost independent of $\omega_\mathrm{i}$.
Then the energy loss of the x ray increases as $\omega_\mathrm{i}$ increases.
In the present RIXS, however, the fluorescencelike behavior would come from a different origin~\cite{Okada2000,Benjamin2014} since these electrons are rather correlated.

The intermediate state of the RIXS process contains a core hole created by an incident photon. Therefore, the Coulomb interaction between the core hole and valence electron ($3d_{x^2-y2}$ electron in cuprates) acting in the intermediate state influences the RIXS spectrum.
It is thus expected that the core-hole Coulomb interaction also affects the $\omega_\mathrm{i}$ dependence of the RIXS spectrum.

In this paper, we perform the Lanczos-type exact diagonalization calculation of the RIXS spectrum for the single-band Hubbard model describing hole- and electron-doped cuprates.
Examining the $\omega_\mathrm{i}$ dependence of non-spin-flip RIXS spectra, we find that the dependence is strongly affected by the value of core-hole Coulomb interaction in hole doping:  A $\omega_\mathrm{i}$-dependent fluorescencelike behavior appears when the core-hole Coulomb interaction is larger than the on-site Coulomb interaction in the Hubbard model, while a $\omega_\mathrm{i}$-independent Raman-like behavior appears for smaller core-hole Coulomb interaction although the distribution of spectral weight depends on $\omega_\mathrm{i}$.
Analyzing main and satellite structures in x-ray absorption (XAS) for hole doping, we find that it is reasonable to take the core-hole Coulomb interaction larger than the on-site Coulomb interaction in modeling RIXS by a single-band model of cuprates.
This suggests a fluorescence-like behavior in RIXS for the $\sigma$-polarized geometry detecting non-spin-flip excitations, being consistent with recent experiments~\cite{Minola2015,Huang2016}.
In such a case, the dynamical charge structure factor is observed through $\sigma$-polarized RIXS by tuning $\omega_\mathrm{i}$ to a satellite structure in XAS for hole doping.

Using the larger core-hole Coulomb interaction, we predict a shift on the high-energy side of the center of spectral distribution in electron doping, which has not yet been observed experimentally.
In the spin-flip channel, main structures exhibit a Raman-like behavior as expected.
In addition, there is a fluorescencelike behavior similar to the non-spin-flip channel, although spectral weight is very small.
The behavior is enhanced with increasing hole carriers.
A detailed experimental work to detect these behaviors is desired in the future. 

This paper is organized as follows.
The Hubbard model and RIXS spectra decomposed into spin-flip and non-spin-flip channels are introduced in Sec.~\ref{Sec2}.
In Sec.~\ref{Sec3}, we calculate the dependence of XAS on the core-hole Coulomb interaction for both hole and electron doping.
The $\omega_\mathrm{i}$ dependence of RIXS spectra are shown in Sec.~\ref{Sec4} and the origin of fluorescencelike and Raman-like behaviors is discussed.
Finally, a summary is given in Sec.~\ref{Sec5}.

\section{Model and Method}
\label{Sec2}
In order to describe $3d$ electrons in the CuO$_2$ plane, we take a single-band Hubbard model given by
\begin{equation}
H_\mathrm{3d}=-t\sum_{i\delta\sigma} c^\dagger_{i\sigma} c_{i+\delta\sigma} -t'\sum_{i\delta'\sigma} c^\dagger_{i\sigma} c_{i+\delta'\sigma} + U\sum_i n_{i\uparrow}n_{i\downarrow},
\label{singleH}
\end{equation}
where $c^\dagger_{i\sigma}$ is the creation operator of an electron with spin $\sigma$ at site $i$, number operator $n_{i\sigma}=c^\dagger_{i\sigma}c_{i\sigma}$, $i+\delta$ ($i+\delta'$) represents the four first (second) nearest-neighbor sites around site $i$, and $t$, $t'$, and $U$ are the nearest-neighbor hopping, the next-nearest-neighbor hopping, and on-site Coulomb interaction, respectively.
We take $U/t=10$ and $t'/t=-0.25$, which are typical values appropriate for cuprates.

Based on the Hubbard model, the XAS for the $L$ edge accompanied by the excitation of an electron from the core Cu$2p$ to Cu$3d$ orbital can be described by
\begin{equation}\label{XAS}
I^\mathrm{XAS}\left(\omega \right) = -\frac{1}{\pi}\mathrm{Im} \left\langle 0 \right| \sum_{l\sigma} c_{l\sigma}\frac{1}{\omega-H_l^j+E_0+i\Gamma} c^\dagger_{l\sigma}\left| 0 \right\rangle,
\end{equation}
where $\left|0 \right\rangle$ represents the ground state with energy $E_0$; $j$ is the total angular momentum of Cu$2p$ with either $j=1/2$ or $j=3/2$; $\Gamma$ is the relaxation time of the core hole; and $H_l^j=H_{3d}-U_\mathrm{c} \sum_\sigma n_{l\sigma} + \varepsilon_j$ with $U_\mathrm{c}$ and $\varepsilon_j$ being the Cu $2p$-$3d$ Coulomb interaction and energy level of Cu $2p$, respectively.
Here, we assume the presence of a Cu$2p$ core hole at site $l$.

In RIXS, tuning polarization of incident and outgoing photons, we can separate excitation with the change of total spin by one ($\Delta S=1$) and excitation with no change of total spin ($\Delta S=0$)~\cite{Haverkort2010,Igarashi2012,Kourtis2012,Tohyama2015}.
We call the former (the latter) spin-flip (non-spin-flip) process hereafter.
The two excitations can be defined as
\begin{eqnarray}\label{IDS0}
I^{\Delta S=0}_\mathbf{q} \left(\Delta \omega \right) &=& \sum\limits_f \left| \left\langle f \right|N^j_\mathbf{q} \left| 0 \right\rangle \right|^2 \delta \left( \Delta \omega  - E_f + E_0 \right),\\
\label{IDS1}
I^{\Delta S=1}_\mathbf{q} \left(\Delta \omega \right) &=& \sum\limits_f \left| \left\langle f \right|S^j_\mathbf{q} \left| 0 \right\rangle \right|^2 \delta \left( \Delta \omega  - E_f + E_0 \right),
\end{eqnarray}
with $S^j_\mathbf{q}=(B^j_{\mathbf{q}\uparrow\uparrow}-B^j_{\mathbf{q}\downarrow\downarrow})/2$, $N^j_\mathbf{q}=B^j_{\mathbf{q}\uparrow\uparrow}+B^j_{\mathbf{q}\downarrow\downarrow}$, and
\begin{equation}\label{Bqw}
B^j_{\mathbf{q}\sigma'\sigma}=\sum_l e^{-i\mathbf{q}\cdot\mathbf{R}_l} c_{l\sigma'}\frac{1}{\omega_\mathrm{i}-H_l^j+E_0+i\Gamma} c^\dagger_{l\sigma},
\end{equation}
where $\left|f \right\rangle$ represents the final state with energy $E_f$; and $\mathbf{R}_l$ is the position vector at site $l$.

When $\Gamma$ is much larger than the remaining terms in the denominator of (\ref{Bqw}), $S^j_\mathbf{q}$ and $N^j_\mathbf{q}$ reduce to $S^z_\mathbf{q}=\sum_l e^{-i\mathbf{q}\cdot\mathbf{R}_l} S^z_l$ and $N_\mathbf{q}=\sum_l e^{-i\mathbf{q}\cdot\mathbf{R}_l} N_l$, respectively, with the $z$ component of the spin operator $S^z_l$ and electron-number operator $N_l$ (the first-collision approximation). In this approximation, (\ref{IDS0}) and (\ref{IDS1}) read the dynamical charge structure factor,
\begin{equation}
N(\mathbf{q},\omega)=\sum\limits_f \left| \left\langle f \right|N_\mathbf{q} \left| 0 \right\rangle \right|^2 \delta \left( \omega  - E_f + E_0 \right), \label{Nqw}
\end{equation}
and the dynamical spin structure factor,
\begin{equation}
S(\mathbf{q},\omega)=\sum\limits_f \left| \left\langle f \right|S^z_\mathbf{q} \left| 0 \right\rangle \right|^2 \delta \left( \omega  - E_f + E_0 \right), \label{Sqw}
\end{equation}
respectively.

When $\Gamma$ is comparable with the remaining terms in (\ref{Bqw}), intersite operators emerge as effective operators of (\ref{Bqw}) in addition to the on-site charge $N_i$ and spin $S_i^z$~\cite{Jia2016}.
In the non-spin-flip process, a two-magnon-type operator is easily expected to contribute to $N^j_\mathbf{q}$.
We thus define the $\mathbf{q}$-dependent dynamical two-magnon correlation function, 
\begin{equation}
M(\mathbf{q},\omega)=\sum\limits_f \left| \left\langle f \right|M_\mathbf{q}^\pm \left| 0 \right\rangle \right|^2 \delta \left( \omega  - E_f + E_0 \right), \label{Mqw}
\end{equation}
with $M_\mathbf{q}^\pm=\sum_\mathbf{k} \left(\cos k_x \pm \cos k_y \right) \mathbf{S}_{\mathbf{k}+\mathbf{q}}\cdot\mathbf{S}_\mathbf{-k}$, where $+$ ($-$) corresponds to A$_{1\mathrm{g}}$ (B$_{1\mathrm{g}}$) representation.

In order to calculate Eqs.~(\ref{IDS0}), (\ref{IDS1}), (\ref{Nqw}), (\ref{Sqw}), and (\ref{Mqw}), we use a Lanczos-type exact diagonalization technique on a $\sqrt{18}\times\sqrt{18}$ cluster under periodic boundary conditions.
We consider hole and electron doping with the carrier concentration $x=n/18$, corresponding to $18-n$ electrons and $18+n$ electrons in the cluster for hole and electron doping, respectively. We set $\Gamma=t$ for both XAS and RIXS.

\section{x-ray absorption spectrum (XAS)}
\label{Sec3}
\begin{figure}
\includegraphics[width=0.45\textwidth]{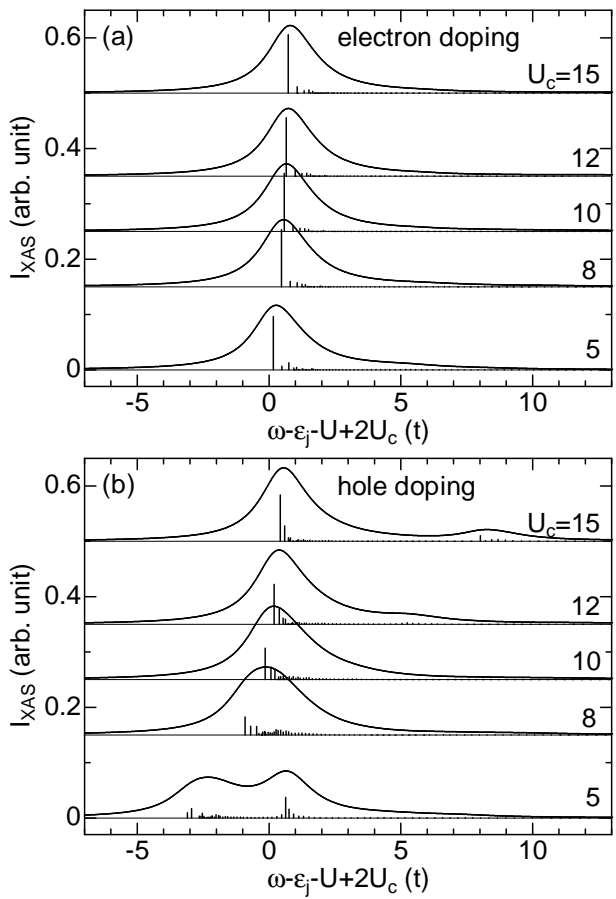}
\caption{The core-hole potential $U_\mathrm{c}$ dependence of XAS spectrum $I^\mathrm{XAS}\left(\omega \right)$ for the $\sqrt{18}\times\sqrt{18}$ periodic cluster of the Hubbard model. (a) Electron doping and (b) hole doping with $x=2/18\sim 0.11$. $U=10t$ and $t'=-0.25t$. The value of $U_\mathrm{c}$ is changed in the range of $U/2\geq U_\mathrm{c} \geq 3U/2$. The solid lines are obtained by a Lorenzian broadening with $\Gamma=t$ for $\delta$ functions shown by bars.}
\label{fig1}
\end{figure}

The value of $U_\mathrm{c}$ used in the literature is ranged from $\sim U/2$~\cite{Jia2016} to $\sim 3U/2$~\cite{Tohyama2015} in describing $L$-edge RIXS for cuprates.
In Fig.~\ref{fig1}, the $U_\mathrm{c}$ dependence of XAS spectrum $I^\mathrm{XAS}\left(\omega \right)$ is shown for $x=0.11$.
In electron doping, the XAS spectra exhibit a single peak independent of $U_\mathrm{c}$, located near $\omega\sim\varepsilon_j+U-2U_\mathrm{c}$.
This is reasonable since kicking an electron from Cu$2p$ to Cu$3d$ creates a doubly occupied state at the core-hole site, similar to half filling as long as $U/2<U_\mathrm{c}$.
This process is denoted as $\tilde{d}^1\rightarrow\underline{\tilde{c}}\tilde{d}^2$, where $\tilde{d}^{1(2)}$ and $\underline{\tilde{c}}$ represent a singly (doubly) occupied state and a core hole, respectively.
We note that an electron carrier cannot contribute to XAS because the carrier already creates the $\tilde{d}^2$ state. 

On the other hand, a clear $U_\mathrm{c}$ dependence appears in hole doping accompanied with a two-peak  structure as shown in Fig.~\ref{fig1}(b).
For $U_\mathrm{c}=15t$ there are two peaks at $\omega-\varepsilon_j-U+2U_\mathrm{c}=0.5t$ and $8t$.
One peak comes from the process $\tilde{d}^1\rightarrow\underline{\tilde{c}}\tilde{d}^2$ same as the case of electron doping and its energy is given by $\varepsilon_j+U-2U_\mathrm{c}$ neglecting the hopping terms.
The other peak is related to a singly occupied state at the core-hole site, where a hole carrier is annihilated by an electron from core Cu$2p$ as denoted by $\tilde{d}^0\rightarrow\underline{\tilde{c}}\tilde{d}^1$ and its peak position is higher in energy by $U_\mathrm{c}-U$ (neglecting the hopping) as compared with the $\underline{\tilde{c}}\tilde{d}^2$ peak.
With decreasing $U_\mathrm{c}$ in Fig.~\ref{fig1}(b), the $\underline{\tilde{c}}\tilde{d}^1$ peak shifts to lower energy and overlaps with the $\underline{\tilde{c}}\tilde{d}^2$ peak near $U_\mathrm{c}\sim U=10t$.
With further decrease, the $\underline{\tilde{c}}\tilde{d}^1$ peak is lower in energy than the $\underline{\tilde{c}}\tilde{d}^2$ peak:  The former peak is at $\omega-\varepsilon_j-U+2U_\mathrm{c}=-2.5t$ and the latter one is at $0.5t$ for $U_\mathrm{c}=5$.
These results suggest that the $U_\mathrm{c}$ dependence of RIXS spectra is strong in hole doping.

The Cu $L_3$-edge XAS in hole-doped cuprates has detected two structures near 930~eV~\cite{Bianconi1987,Sarma1988}: One is a large peak coming from the process of $3d^9\rightarrow\underline{2p}3d^{10}$ ($\underline{2p}=$ Cu$2p$ core hole) and the other is a weak structure related to carrier concentration~\cite{Ghigna1998} given by the process of  $3d^9\underline{L}\rightarrow\underline{2p}3d^{10}\underline{L}$ ($\underline{L}=$ oxygen ligand hole).
The latter structure is higher by roughly 1.5~eV in energy than the former one~\cite{Bianconi1987,Sarma1988,Ghigna1998}.
By using Cu-O clusters with appropriate parameters, the two structures have been obtained in hole doping~\cite{Veenendaal1994}.
A two-band CuO model used in RIXS calculations~\cite{Tohyama2015} also reproduced a two-structure behavior similar to the experiments (not shown).

In comparing the experimental data with the XAS in the single-band Hubbard model, it would be reasonable to assign $3d^9\rightarrow\underline{2p}3d^{10}$ ($3d^9\underline{L}\rightarrow\underline{2p}3d^{10}\underline{L}$) to $\tilde{d}^1\rightarrow\underline{\tilde{c}}\tilde{d}^2$ ($\tilde{d}^0\rightarrow\underline{\tilde{c}}\tilde{d}^1$) by taking into account the correspondence between the Zhang-Rice singlet state ($3d^9\underline{L}$) and the lower-Hubbard-band state ($\tilde{d}^0$).
Taking a standard value of $t=0.35$~eV, we can find that the energy separation of the two structures in hole-doped XAS shown in Fig.~\ref{fig1}(b) becomes an experimental value (1.5~eV) when $U_\mathrm{c}\sim 12t$.
This means that $U_\mathrm{c}>U=10t$.
However, a smaller value of $U_\mathrm{c}$ satisfying $U_\mathrm{c}<U$ has been used in the literature~\cite{Huang2016,Jia2016}.
Therefore, it is interesting to examine the effect of $U_\mathrm{c}$ on the $\omega_\mathrm{i}$ dependence of the RIXS spectrum.
In the following, keeping $U=10t$, we use $U_\mathrm{c}=12t$ and $U_\mathrm{c}=8t$ as a representative parameter for $U_\mathrm{c}>U$ and $U_\mathrm{c}<U$, respectively.

\section{RIXS spectrum}
\label{Sec4}
\subsection{Non-spin-flip channel}
\label{Sec4.1}
\subsubsection{Hole doping}
\begin{figure}
\includegraphics[width=0.45\textwidth]{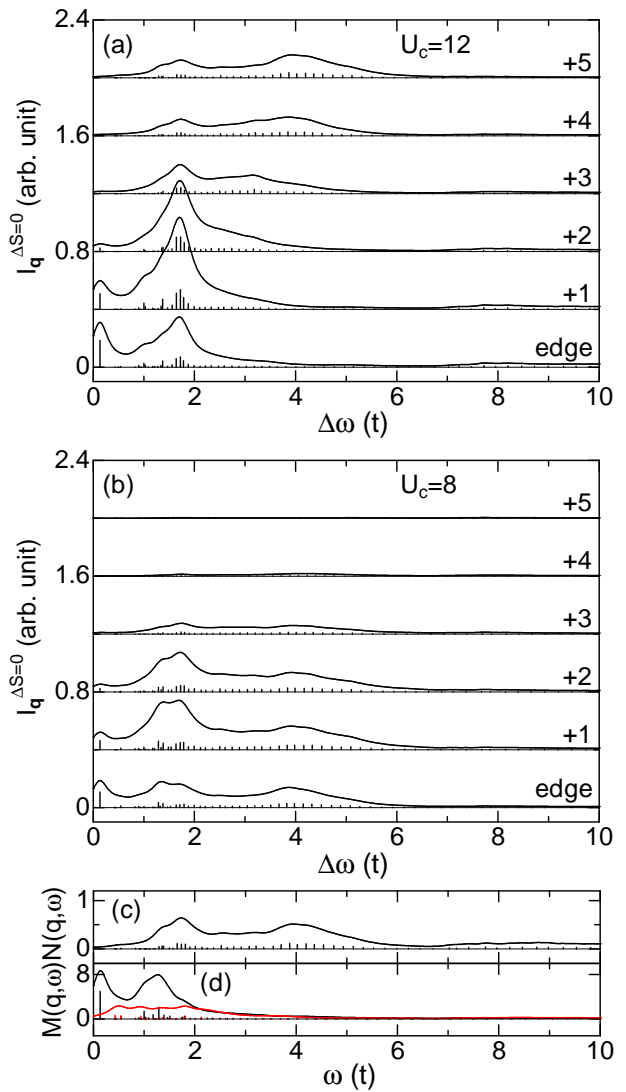}
\caption{
(Color online) The incident-phonon energy $\omega_\mathrm{i} $ dependence of the non-spin-flip spectrum $I^{\Delta S=0}_\mathbf{q}$ at $\mathbf{q}=(2\pi/3,0)$ for the hole-doped $\sqrt{18}\times\sqrt{18}$ Hubbard cluster with $x=2/18\sim 0.11$.
(a) $U_\mathrm{c}=12t$ and (b) $U_\mathrm{c}=8t$.
Parameters are $U=10t$, $t'=-0.25t$, and $\Gamma=t$.
$\omega_\mathrm{i}$ at the bottom panel is set to the edge of the XAS spectrum and increases by $t$ from bottom to top.
(c) The dynamical charge structure factor $N(\mathbf{q},\omega)$ at $\mathbf{q}=(2\pi/3,0)$.
(d) The $\mathbf{q}$-dependent dynamical two-magnon correlation function $M(\mathbf{q},\omega)$ at $\mathbf{q}=(2\pi/3,0)$. Black (red) line, B$_{1\mathrm{g}}$ (A$_{1\mathrm{g}}$) mode.
}
\label{fig2}
\end{figure}

The incident-photon energy $\omega_\mathrm{i}$ dependence of non-spin-flip intensity $I^{\Delta S=0}_\mathbf{q}$ at $\mathbf{q}=(2\pi/3,0)$ in hole doping with $x=0.11$ is shown for the case of $U_\mathrm{c}=12t>U$ in Fig.~\ref{fig2}(a).
The bottom spectrum exhibits the spectrum obtained by tuning $\omega_\mathrm{i}$ at the edge of the lowest-energy peak in the XAS spectrum shown in Fig.~\ref{fig1}(b).
Here, the edge is defined as the lowest-energy eigenstate represented by the lowest-energy vertical bar in Fig.~\ref{fig1}(b).
There are two peaks at $\omega=1.8t$ and $\omega=0.2t$ together with a broad tail around $\omega=3t$.
The peak at $\omega=1.8t$ is similar to that at half filling, which is originated from two-magnon excitation.
We note that in this energy region  there are also charge excitations coming from hole carrier as seen from the dynamical charge structure factor $N(\mathbf{q},\omega)$ in Fig.~\ref{fig2}(c).
On the other hand, the peak at $\omega=0.2t$ does not have a corresponding structure in   $N(\mathbf{q},\omega)$, indicating that the peak is related to an excitation beyond a simple two-particle response~\cite{Jia2016}.
Actually, we find that the peak at $\omega=0.2t$ appears in $M(\mathbf{q},\omega)$ with B$_{1\mathrm{g}}$, as shown in Fig.~\ref{fig2}(d).
The broad tail around $\omega=3t$ may partly come from charge excitations, but spectral shape is different from $N(\mathbf{q},\omega)$.
This is reasonable since the edge of XAS is mainly composed of the $\underline{\tilde{c}}\tilde{d}^2$ state to which hole carriers do not contribute.

\begin{figure}
\includegraphics[width=0.45\textwidth]{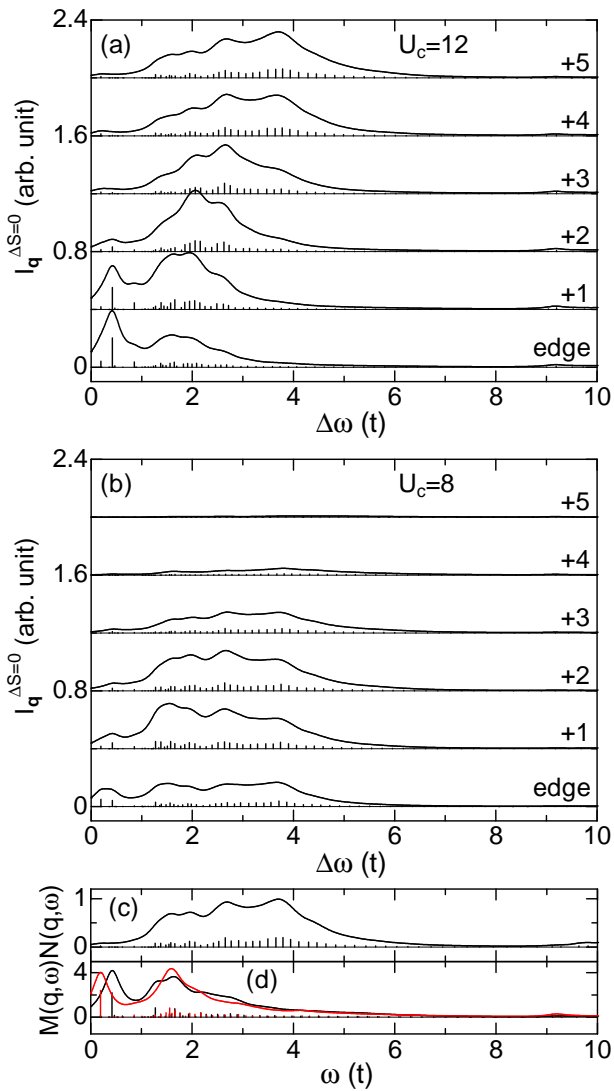}
\caption{(Color online) Same as Fig.~\ref{fig2} but $x=4/18\sim 0.22$.}
\label{fig3}
\end{figure}

When $\omega_\mathrm{i}$ is increased by $t$ from the edge, $I^{\Delta S=0}_\mathbf{q}$ increases as shown in Fig.~\ref{fig2}(a).
This is simply due to the fact that the peak of the XAS is located slightly above the edge position and thus resonance becomes maximum when $\omega_\mathrm{i}$ is larger than the edge energy.
With further increasing $\omega_\mathrm{i}$, spectral weight gradually decreases as the resonance condition becomes weaker.
Furthermore, the distribution of spectral weight becomes wider, accompanied by the reduction of peak heights.
When $\omega_\mathrm{i}$ is higher than the edge position by $5t$, where the incident photon resonates to a satellite structure in XAS, spectral distribution becomes similar to $N(\mathbf{q},\omega)$.
This is again reasonable since the incident photon resonates mainly to the $\underline{\tilde{c}}\tilde{d}^1$ state [see Fig.~\ref{fig1}(b)] and thus hole carriers contribute to the RIXS process through the intermediate state.
As a consequence, a characteristic structure in the spectral weight gradually changes from the low-energy peak near $\Delta\omega=2t$ to the high-energy broad peak near $\Delta\omega=4t$ with increasing $\omega_\mathrm{i}$.
This evolution of $I^{\Delta S=0}_\mathbf{q}$ with $\omega_\mathrm{i}$ exhibits a fluorescencelike $\omega_\mathrm{i}$ dependence:  The energy position of main spectral weight increases with $\omega_\mathrm{i}$.
Such dependence is due to the fact that particle-hole intraband excitations can couple to the $\underline{\tilde{c}}\tilde{d}^1$ state extended above the absorption edge in XAS. 
Note that the amount of the energy shift is not the same as the increase of $\omega_\mathrm{i}$, since a photoexcited electron is no longer independent of other electrons unlike the case of the normal XES.

In contrast to the $U_\mathrm{c}=12t$ case, the $\omega_\mathrm{i}$ dependence for $U_\mathrm{c}=8t<U$ shows less dispersive spectral distribution as seen in Fig.~\ref{fig2}(b). When $\omega_\mathrm{i}$ is tuned to the edge of XAS, there is a peak structure at $\omega=4t$, which is originated from the charge excitation shown in Fig.~\ref{fig2}(c).
This structure remains with increasing $\omega_\mathrm{i}$, resulting in a Raman-like $\omega_\mathrm{i}$ dependence.
This behavior in sharp contrast with the $U_\mathrm{c}=12t$ case can be attributed to the difference of intermediate states.
The main peak in XAS [Fig.~\ref{fig1}(b)] shows a similar single-peak structure for both the $U_\mathrm{c}=8t$ and $U_\mathrm{c}=12t$ cases.
However, in contrast to the $U_\mathrm{c}=12t$ case, not only a doubly occupied state $\underline{\tilde{c}}\tilde{d}^2$ but also a singly occupied state $\underline{\tilde{c}}\tilde{d}^1$ contributes to the main peak for $U_\mathrm{c}=8t$ as discussed above.
The singly occupied state can couple to particle-hole excitations in the final state of RIXS, resulting in a wide energy spectral distribution independent of the value of $\omega_\mathrm{i}$.

The fluorescencelike $\omega_\mathrm{i}$ dependence at $U_\mathrm{c}=12t$ becomes more clear when hole carriers are increased.
Figure~\ref{fig3} shows $I^{\Delta S=0}_\mathbf{q}$, $N(\mathbf{q},\omega)$, and $M(\mathbf{q},\omega)$ at $\mathbf{q}=(2\pi/3,0)$ for $x=0.22$.
As for $x=0.11$, low-energy excitations below $\Delta\omega\sim2t$ in $I_\mathbf{q}^{\Delta S=0}$ at the absorption edge [see Figs.~\ref{fig3}(a) and \ref{fig3}(b)] are partly contributed by two-magnon excitations as expected from $M(\mathbf{q},\omega)$ shown in Fig.~\ref{fig3}(d).
It is clear in Fig.~\ref{fig3}(a) that, with increasing $\omega_\mathrm{i}$ from the edge position, the spectral distribution smoothly shifts to the high-energy side accompanied by the reduction of low-energy weight similar to the $x=0.11$ case and eventually resembles $N(\mathbf{q},\omega)$ shown in Fig.~\ref{fig3}(c).
This smooth shift indicates that the fluorescence-like behavior can be seen more clearly in the overdoped region.
On the other hand, $\omega_\mathrm{i}$ dependence at $U_\mathrm{c}=8t$ exhibits a Raman-like behavior consistent with the case of $x=0.11$. 

We note that these $\omega_\mathrm{i}$ dependencies mentioned above are seen at not only $\mathbf{q}=(2\pi/3,0)$ but also $\mathbf{q}=(\pi/3,\pi/3)$ defined in the $\sqrt{18}\times\sqrt{18}$ cluster (not shown).

The $\omega_\mathrm{i}$ dependence of the RIXS spectrum has been measured in YBa$_2$Cu$_3$O$_{6+x}$~\cite{Minola2015}.
In the experiment, the $\sigma$ polarization of the incident photon approximately correspond to non-spin-flip spectrum $I^{\Delta S=0}_\mathbf{q}$.
Comparing the calculated $I^{\Delta S=0}_{\mathbf{q}=(2\pi/3,0)}$ in Figs.~\ref{fig2} and \ref{fig3} with the experimental data at $\mathbf{q}=(0.74\pi,0)$ for the overdoped sample~\cite{Minola2015}, we find that the $U_\mathrm{c}=12t$ case corresponds to the experiment, since there is a fluorescencelike  behavior.
Encouraged by the correspondence, we use $U_\mathrm{c}=12t$ in the following as a representative value for cuprates.

\begin{figure}
\includegraphics[width=0.45\textwidth]{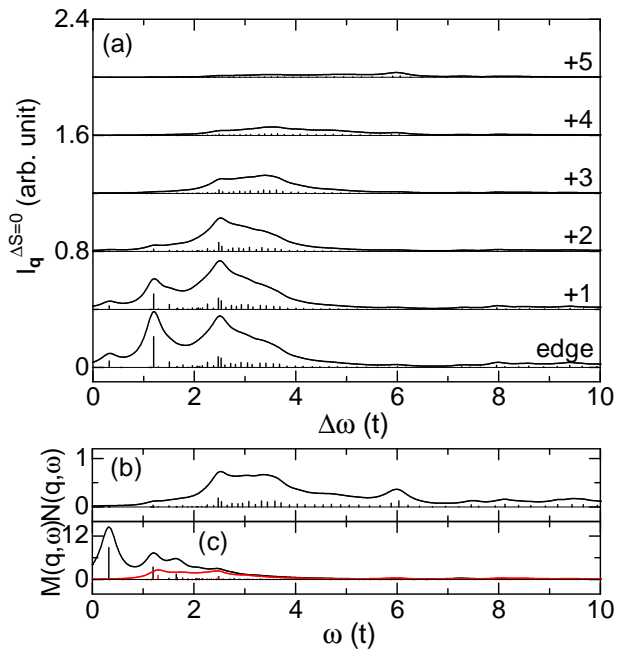}
\caption{
(Color online) (a) The incident-phonon energy $\omega_\mathrm{i} $ dependence of the non-spin-flip spectrum $I^{\Delta S=0}_\mathbf{q}$ at $\mathbf{q}=(2\pi/3,0)$ for the electron-doped $\sqrt{18}\times\sqrt{18}$ Hubbard cluster with $x=2/18\sim 0.11$.
Parameters are $U=10t$, $t'=-0.25t$, $U_\mathrm{c}=12t$, and $\Gamma=t$.
$\omega_\mathrm{i}$ at the bottom panel is set to the edge of the XAS spectrum and increases by $t$ from bottom to top.
(b) The dynamical charge structure factor $N(\mathbf{q},\omega)$ at $\mathbf{q}=(2\pi/3,0)$.
(c) The $\mathbf{q}$-dependent dynamical two-magnon correlation function $M(\mathbf{q},\omega)$ at $\mathbf{q}=(2\pi/3,0)$.
Black (red) line, B$_{1\mathrm{g}}$ (A$_{1\mathrm{g}}$) mode.
}
\label{fig4}
\end{figure}

\subsubsection{Electron doping}
Figure~\ref{fig4} shows the $\omega_\mathrm{i}$ dependence of the non-spin-flip spectrum $I^{\Delta S=0}_\mathbf{q}$ at $\mathbf{q}=(2\pi/3,0)$ for electron doping.
When $\omega_\mathrm{i}$ is tuned to the edge, there are two main structures at $\Delta\omega=1.2t$ and $2.5t$.
The latter is partly contributed from charge excitations due to electron carriers as expected from $N(\mathbf{q},\omega)$ shown in Fig.~\ref{fig4}(b).
In $N(\mathbf{q},\omega)$, however, there is no corresponding structure for the former structure, indicating a possible two-magnon origin.
In fact, $M(\mathbf{q},\omega)$ shown in Fig.~\ref{fig4}(c) exhibits an enhancement at the former energy.
We further notice that there is no prominent structure at $\omega=6t$ where $N(\mathbf{q},\omega)$ shows a peak structure. 

With increasing $\omega_\mathrm{i}$, the two main structures gradually lose their weight and a broad structure near $\Delta\omega=3t$ remains accompanied with an small enhancement around $\Delta\omega=6t$.
The spectral distribution, for example, for $\omega_\mathrm{i}$ higher by $\sim3t$ from the edge, resembles $N(\mathbf{q},\omega)$.
With further increasing $\omega_\mathrm{i}$, the center of spectral distribution shifts to the high-energy side, although the spectral weight is strongly reduced as compared with the case of hole doping.

We note that the origin of such a fluorescencelike behavior is different from the case of hole doping.
In hole doping, the spectral weight similar to $N(\mathbf{q},\omega)$ at high $\omega_\mathrm{i}$ is constructed by hole carriers through the $\underline{\tilde{c}}\tilde{d}^1$ state.
On the other hand, in electron doping there is no direct contribution from electron carriers in the intermediate state in RIXS, since the carriers already make an inactive $\tilde{d}^2$ states for XAS.
Therefore, the spectral weight similar to $N(\mathbf{q},\omega)$ in electron doping may be due to an indirect effect of electron carriers appearing in the XAS spectra as a broad structure with small spectral weight above the main peak as seen in Fig.~\ref{fig1}(a).
This is the reason for weak fluorescencelike intensity in electron doping.
Detailed examinations of $\omega_\mathrm{i}$ dependence in the $\sigma$-polarized RIXS is desired to confirm theoretical predictions for electron doping.

\begin{figure}
\includegraphics[width=0.45\textwidth]{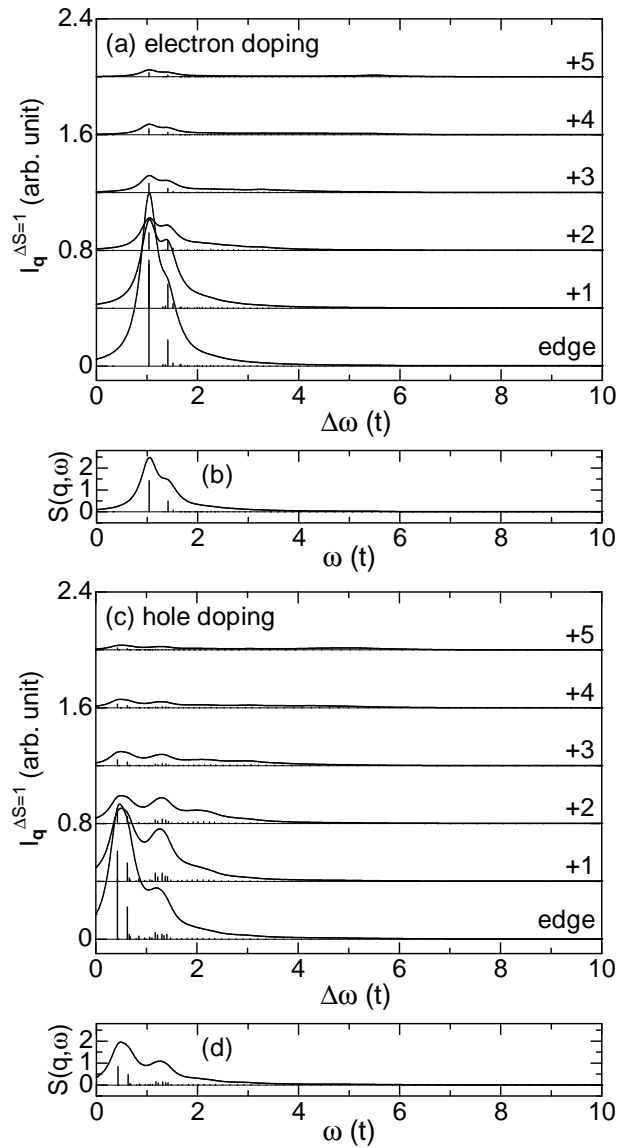}
\caption{
The incident-phonon energy $\omega_\mathrm{i} $ dependence of the spin-flip spectrum $I^{\Delta S=1}_\mathbf{q}$ at $\mathbf{q}=(2\pi/3,0)$ for the $\sqrt{18}\times\sqrt{18}$ Hubbard cluster with $x=2/18\sim 0.11$ for (a) electron doping and (c) hole doping.
Parameters are $U=10t$, $t'=-0.25t$, $U_\mathrm{c}=15t$, and $\Gamma=t$.
$\omega_\mathrm{i}$ at the bottom panel is set to the edge of the XAS spectrum and increases by $t$ from bottom to top.
The dynamical spin structure factor $S(\mathbf{q},\omega)$ at $\mathbf{q}=(2\pi/3,0)$ for (b) electron doping and (d) hole doping.
}
\label{fig5}
\end{figure}

\begin{figure}
\includegraphics[width=0.46\textwidth]{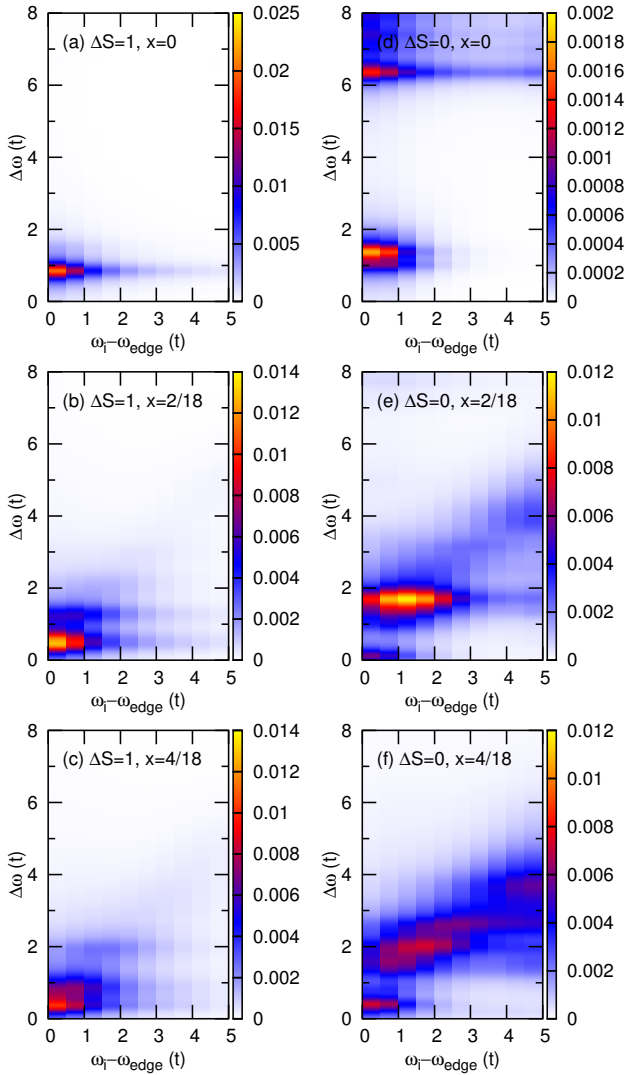}
\caption{
(Color online) The contour plot of incident-phonon energy $\omega_\mathrm{i} $ dependence of spin-flip spectra $I^{\Delta S=1}_\mathbf{q}$ and non-spin-flip spectra $I^{\Delta S=0}_\mathbf{q}$ at $\mathbf{q}=(2\pi/3,0)$ for the hole-doped $\sqrt{18}\times\sqrt{18}$ Hubbard cluster with parameters $U=10t$, $t'=-0.25t$, $U_\mathrm{c}=12t$ and $\Gamma=t$.
(a) $x=0$, (b) $x=2/18\sim 0.11$, and (c) $x=4/18\sim 0.22$ for $I^{\Delta S=1}_\mathbf{q}$.
(d) $x=0$, (e) $x\sim 0.11$, and (f) $x\sim 0.22$ for $I^{\Delta S=0}_\mathbf{q}$.
}
\label{fig6}
\end{figure}

\subsection{Spin-flip channel}
\label{Sec4.2}
Since particle-hole excitations are involved in not only the non-spin-flip channel but also the spin-flip one, it is interesting to examine how the spin-flip spectrum $I^{\Delta S=1}_\mathbf{q}$ is dependent on $\omega_\mathrm{i}$.
Figure~\ref{fig5} shows $I^{\Delta S=1}_\mathbf{q}$ at $\mathbf{q}=(2\pi/3,0)$ for both electron and hole dopings.
Comparing Figs.~\ref{fig5}(a) and \ref{fig5}(b) for electron doping and Figs.~\ref{fig5}(c) and \ref{fig5}(d) for hole doping, we find that $I^{\Delta S=1}_\mathbf{q}$ for $\omega_\mathrm{i}$ tuned to the absorption edge resembles $S(\mathbf{q},\omega)$.
This is in contrast to the case of $I^{\Delta S=0}_\mathbf{q}$ as discussed in Sec.~\ref{Sec4.1}~\cite{Jia2016}.
With increasing $\omega_\mathrm{i}$ from the edge, low-energy spin-flip excitations below $\omega=2t$ in both dopings remain showing a Raman-like behavior.
In addition, there is a dispersive fluorescencelike structure, which is similar to the case of the non-spin-flip spectrum $I^{\Delta S=0}_\mathbf{q}$, although the intensity is very small above $\Delta\omega=2t$.
This indicates the presence of particle-hole excitations in the spin-flip channel.

In order to make clear the effect of hole carriers on the $\omega_\mathrm{i}$ dependence, we show the contour map of $I^{\Delta S=1}_\mathbf{q}$ at $\mathbf{q}=(2\pi/3,0)$ together with the non-spin flip $I^{\Delta S=0}_\mathbf{q}$ for the realistic value of $U_\mathrm{c}=12t$ in Fig.~\ref{fig6}.
At half filling $x=0$ [Figs.~\ref{fig6}(a) and \ref{fig6}(d)], both spectra exhibit Raman-like $\omega_\mathrm{i}$ dependence.
$I^{\Delta S=0}_\mathbf{q}$ clearly shows fluorescencelike $\omega_\mathrm{i}$ dependence with increasing $x$ as expected.
The spin-flip spectra in Figs.~\ref{fig6}(b) and ~\ref{fig6}(c) also show fluorescencelike $\omega_\mathrm{i}$ dependence in the same $\Delta\omega$ region as $I^{\Delta S=0}_\mathbf{q}$, though its intensity is weak as compared with low-energy excitations.
We can find that the fluorescencelike intensity increases with increasing $x$, being consistent with the view that particle-hole excitations contribute to RIXS spectra in the spin-flip channel when carriers become more itinerant.
A detailed examination of $\omega_\mathrm{i}$ dependence of RIXS for the $\pi$-polarized incident photon and different carrier concentrations is desired in the near future.

\section{Summary}
\label{Sec5}
Examining the $\omega_\mathrm{i}$ dependence of non-spin-flip RIXS spectra by the exact diagonalization calculation  for the single-band Hubbard model, we have found that the dependence is strongly affected by the value of the core-hole Coulomb interaction in hole doping:  A fluorescencelike behavior appears when the core-hole Coulomb interaction $U_\mathrm{c}$ is larger than the on-site Coulomb interaction $U$ in the Hubbard model ($U_\mathrm{c}>U$), while a Raman-like behavior appears for $U_\mathrm{c}<U$, although the distribution of spectral weight depends on $\omega_\mathrm{i}$.
Comparing the calculated energy separation between a main peak and a satellite structure in XAS with the corresponding experimental data, we have confirmed that $U_\mathrm{c}>U$ for the single-band Hubbard model, suggesting a fluorescencelike behavior in RIXS for the $\sigma$-polarized geometry detecting non-spin-flip excitations.
This is consistent with recent experimental observations for overdoped YBa$_2$Cu$_3$O$_{6+x}$~\cite{Minola2015}.
We predict that the dynamical charge structure factor is observed through RIXS by tuning $\omega_\mathrm{i}$ to the satellite structure.
Using the same $U_\mathrm{c}$ value, we predict a shift on the high-energy side of the center of spectral distribution in electron doping, whose intensity is reduced as compared with the case of hole doping.
In the spin-flip channel, main structures exhibit a Raman-like behavior as expected but there is  a fluorescencelike behavior similar to the non-spin-flip case, although the spectral weight is very small.
A detailed experimental work to detect these behaviors is highly desired in the near future. 

\begin{acknowledgments}
This work was supported by the Japan Society for the Promotion of Science, KAKENHI (Grants No. 26287079, 15H03553 and 16H04004) and by HPCI Strategic Programs for Innovative Research (Grants No. hp140078) and Computational Materials Science Initiative from Ministry of Education, Culture, Sports, Science, and Technology.
\end{acknowledgments}

\nocite{*}


\end{document}